\documentclass[aps, 10pt, prl, amsmath, longbibliography, twocolumn, superscriptaddress, floatfix]{revtex4-1}
\usepackage[plainpages=false,colorlinks=true,linkcolor=blue, citecolor=blue, urlcolor=blue]{hyperref}
\usepackage[english]{babel}
\usepackage{graphicx}
\usepackage[charter]{mathdesign}
\usepackage{bm}
\usepackage[svgnames]{xcolor}

\newcommand\gv{\mathbf{g}}

\newcommand\xv{\mathbf{x}}
\newcommand\yv{\mathbf{y}}

\newcommand\rv{\mathbf{r}}
\newcommand\tv{\mathbf{t}}
\newcommand\kv{\mathbf{k}}

\newcommand\Kv{\mathbf{K}}
\newcommand\Gv{\mathbf{G}} 

\newcommand\Sv{\mathbf{S}}
\newcommand\kvt{\mathbf{\tilde k}}
\renewcommand\a{\alpha}
\renewcommand\b{\beta}
\newcommand\s{\sigma}

\newcommand\Sigmav{\bm{\Sigma}}
\newcommand\epsilonv{\bm{\epsilon}}
\newcommand\thetav{\bm{\theta}}
\newcommand\Gammav{\bm{\Gamma}}
\newcommand{\dg}{\dagger}
\newcommand{\fdg}{{\phantom{\dagger}}}
\newcommand{\up}{{\uparrow}}
\newcommand{\down}{\downarrow}

\newcommand\Tr{\operatorname{Tr}}
\renewcommand\Im{\operatorname{Im}}

\begin{document}
\title{Spin texture in a bilayer high-temperature cuprate superconductor}
\author{Xiancong Lu}
\affiliation{Department of Physics, Xiamen University, Xiamen 361005, China}
\author{D. S\'en\'echal}
\affiliation{D\'epartement de physique and Institut quantique, Universit\'e de Sherbrooke, Sherbrooke, Qu\'ebec, Canada J1K 2R1}

\begin{abstract}
We investigate the possibility of spin texture in the bilayer cuprate superconductor $\rm Bi_2Sr_2CaCu_2O_{8+\delta}$ using cluster dynamical mean field theory (CDMFT). 
The one-band Hubbard model with a small interlayer hopping and a Rashba spin-orbit coupling is used to describe the material.
The $d$-wave order parameter is not much affected by the presence of the Rashba coupling, but a small triplet component appears.
We find a spin texture circulating in the same direction around $\kv=(0,0)$ and $\kv=(\pi,\pi)$ and stable against the superconducting phase. The amplitude of the spin structure, however, is strongly affected by the pseudogap phenomenon, more so than the spectral function itself.
\end{abstract}
\maketitle

While electron-electron interactions are a key ingredient in the study of quantum materials, the presence of a spin-orbit coupling (SOC) is the source of new emergent phenomena, especially in heavy transition metal compounds~\cite{wi.ch.14}.
The SOC is a key ingredient of the topological states of matter~\cite{ha.ka.10,qi.zh.11}.
The interplay or competition between SOC and electron correlations is relevant in systems like the heavy fermion superlattices~\cite{sh.go.16}, iridium oxides~\cite{ra.le.16}, and optical lattices~\cite{ma.ko.15}, in which exotic phases are expected. 
Within the Rashba-Hubbard model, a mixed singlet-triplet superconducting state~\cite{fr.ag.04,yo.on.07,ya.si.08,ta.ka.08,gr.sc.18,lu.se.18,gh.ka.19,no.ya.20,wo.ra.20},
novel magnetism~\cite{zh.wu.15,gr.be.20}, and nontrivial topological properties~\cite{fa.pe.14,la.re.14,lu.se.18,marc.17} are theoretically predicted.

The spin-orbit coupling also manifests itself in a globally centrosymmetric crystal, which contains subunits in which the
inversion symmetry is broken locally~\cite{fi.lo.11,ma.si.12,si.ag.14,zh.li.14}, \textit{i.e.}, a locally non-centrosymmetric crystal.  
The bilayer SOC system is a typical example, in which the SOC on two non-equivalent layers have opposite signs, such as the $\rm CeCoIn_5/YbCoIn_5$ hybrid structure \cite{go.mi.12}, $\rm  SrPtAs$ \cite{ni.ku.11,si.ag.14}, and bilayer transition metal dichalcogenides (TMDs)~\cite{jo.yu.14,liucx.17}. 
The absence of local inversion symmetry in the bilayer system can lead to a ``hidden'' spin polarization~\cite{zh.li.14,ri.ma.14}, nontrivial topological states~\cite{na.ta.12, da.ba.13,do.wa.15}, and unconventional superconductivity~\cite{si.ag.14,liucx.17,is.ya.18}.

Recent spin- and angle-resolved photoemission spectroscopy (SARPES) experiments have shown that, in one of the most studied cuprate superconductors $\rm Bi_2Sr_2CaCu_2O_{8+\delta}$ (Bi2212), a striking spin texture develops in the Brillouin zone with spin-momentum locking~\cite{go.li.18}. 
The observed spin texture is consistent with the prediction of a bilayer model with opposite Rashba SOC on the two CuO layers of the unit cell. 
This has motivated new studies focusing on the hidden SOC in high-$T_c$ cuprates \cite{ra.al.19,hi.ya.19,atki.20}. 
However, correlation effects and a possible competition between the spin texture and $d$-wave superconductivity have not been fully considered so far.
In this paper, we will address these important issues in a fully dynamical study employing cluster dynamical mean field theory (CDMFT).
	
\paragraph*{Model ---}

To describe the locally non-centrosymmetric bilayer high-$T_c$ cuprate, we use the following tight-binding bilayer Rashba model \cite{go.li.18,ra.al.19,hi.ya.19,ha.ra.15}:
\begin{eqnarray}
H = H_{\rm kin} + H_{\rm SOC} + H_{\perp} + H_U. 
\label{eq:H}
\end{eqnarray}
The non-interacting part includes three terms,
\begin{eqnarray}
H_{\rm kin} &=& \sum_{l,\kv,\s} \varepsilon(\kv) c_{l\kv\s}^\dg
            c_{l\kv\s}^\fdg \label{eq:H0a}\\
H_{\rm SOC} &=& \sum_{l,\kv,\s,\s'} \lambda_l \gv_\kv \cdot
            \boldsymbol{\s}^{\s\s'}  
            c_{l\kv\s}^\dg c_{l\kv\s'}^\fdg \label{eq:H0b} \\ 
H_{\perp} &=& \sum_{\kv,\s} t_{\perp}(\kv) \Big[ 
              c_{1\kv\s}^\dg c_{2\kv\s}^\fdg + \mathrm{H.c} \Big] \label{eq:H0c}
\end{eqnarray}

where $c_{l\kv\s}(c_{l\kv\s}^\dg)$ is the annihilation (creation) operator of an electron on the $l$th layer ($l=1,2$) with spin $\s=\up,\down$ and wave vector $\kv$. 
The dispersion relation on a square lattice is $\varepsilon(\kv) = -2t(\cos k_x + \cos k_y) +4t'\cos k_x\cos k_y-2t''(\cos 2k_x + \cos 2k_y) - \mu $, in which the hopping terms up to third nearest neighbor ($t$, $t'$, and $t''$) and the chemical potential $\mu$ are included. 
$\gv_\kv=(-\sin k_y, \sin k_x, 0)$ defines the antisymmetric SOC of Rashba type and $\boldsymbol{\s}$ is the vector of Pauli matrices. 
Due to the global inversion symmetry, the SOC in layers 1 ad 2 are opposite in sign: $\lambda_1=-\lambda_2$. 
The interlayer hoping is $t_{\perp}(\kv)=t_z(\cos k_x - \cos k_y)^2$, which causes the bilayer splitting in high-$T_c$ cuprate. 
For Bi2212, the nearest neighbor hopping is $t=360$~meV \cite{ma.sa.05,dr.pl.18}. 
In the remainder of this paper, we set $t=1$ as the energy unit, and choose the other tight-binding parameters to be $t'=-0.3$, $t''=0.1$ and $t_z=0.08$ \cite{dr.pl.18}. Finally, we set $\lambda_1=-\lambda_2=0.06$.

The interacting part of Hamiltonian reads
\begin{eqnarray}
H_U = U\sum_{\rv,l} n_{\rv,l,\up} n_{\rv,l,\down} 
\end{eqnarray}
in which $n_{\rv,l,\s}$ is the number of electrons of spin $\s$ at lattice site $\rv$ of layer $l$. Only on-site interactions are considered.
		
\paragraph*{CDMFT  ---}

In order to reveal the spin texture arising in model~\eqref{eq:H}, we use cluster dynamical mean-field theory
(CDMFT)~\cite{li.ka.00,ko.sa.01,li.is.08,sene.15} with an exact diagonalization solver at zero temperature (or ED-CDMFT).  
In this approach, the infinite lattice is tiled into identical units (or \textit{supercells}) defining a superlattice. 
The supercell is made of one or more \textit{clusters}, each of which coupled to a bath of uncorrelated, auxiliary orbitals.  
The parameters describing this bath (energy levels, hybridization, etc.)  are then found by imposing a self-consistency
condition.

\begin{figure}[h]\centering
$$\mathbf{(a)}\vcenter{\hbox{\includegraphics[width=0.5\columnwidth]{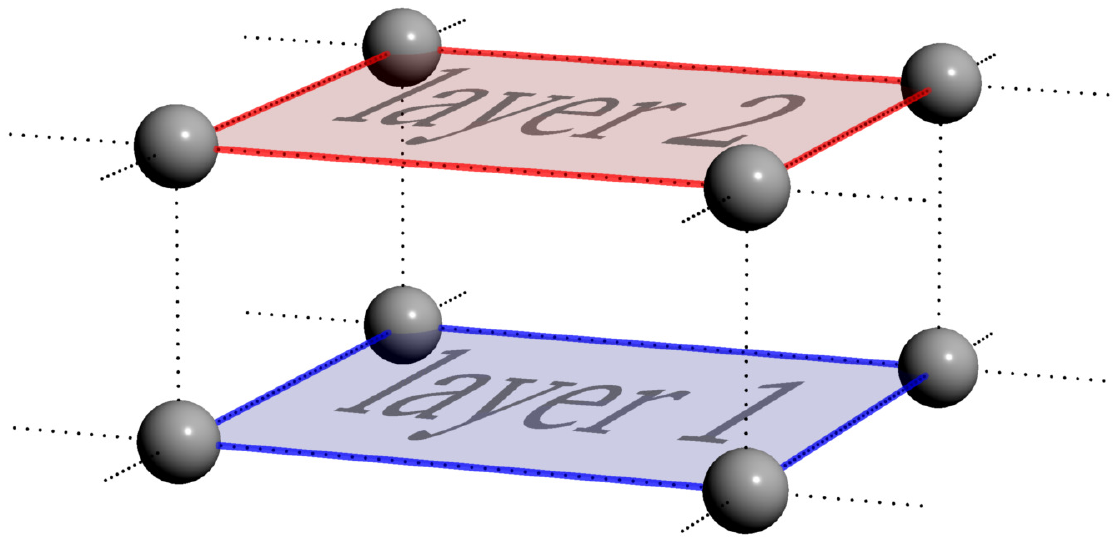}}}$$
$$\mathbf{(b)}\vcenter{\hbox{\includegraphics[width=0.7\columnwidth]{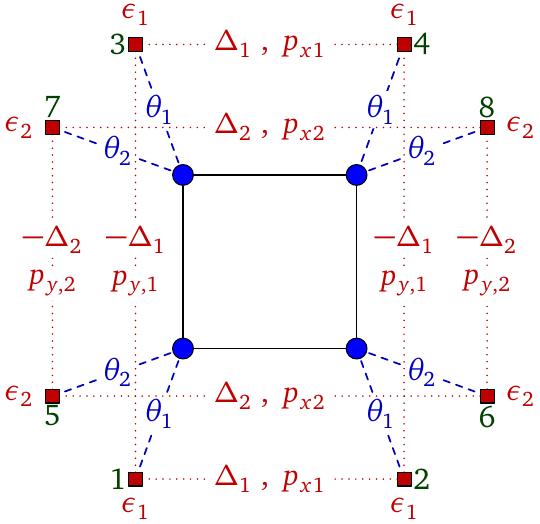}}}$$
\caption{(Color online). The cluster-bath systems used in our implementation of ED-CDMFT. 
Panel (a) : position of the two clusters forming the supercell in 3D.
Panel (b) : the cluster (blue) and bath (red) orbitals, with the various bath parameters
used in this study. See text for details.\label{fig:cluster_bath}}
\end{figure}

In this work the supercell is made of two superimposed, four-site plaquettes (one per layer), each of which coupled to a bath of eight uncorrelated orbitals. 
The cluster-bath system, or \textit{impurity model}, is illustrated on Fig.~\ref{fig:cluster_bath} and defined by the following
Anderson impurity model (AIM):
\begin{equation}\label{eq:Himp}
H_{\rm imp} = H_c + \sum_{\mu,\a} \theta_{\mu,\a} \left(c_\mu^\dg a_\a^\fdg + \mbox{H.c.} \right)
+ \sum_{\a\b} \epsilon_{\a\b} a_\a^\dg a_\b^\fdg~~,
\end{equation}
where $H_c$ is the Hamiltonian~\eqref{eq:H}, but restricted to a single cluster, and $c_\mu$ and $a_\a$ destroy electrons on the cluster sites and the bath orbitals, respectively.
Probing superconductivity forces us to use the Nambu formalism, in which each degree of freedom is occurring in particle and hole form in a multiplet. 
Thus, the index $\mu$ is a composite index comprising cluster site $i$, spin and Nambu indices: $c_i = (c^\fdg_{i\up}, c^\fdg_{i\down}, c^\dg_{i\up}, c^\dg_{i\down})$. This index takes $4\times4=16$ values in the particular AIM that we use.
Likewise, the index $\a$ comprises bath orbital index $r$, spin and Nambu indices and takes $4\times8=32$ values:
$a_r = (a^\fdg_{r\up}, a^\fdg_{r\down}, a^\dg_{r\up}, a^\dg_{r\down})$.
$\theta_{\mu\a}$ is a complex-valued, $16\times32$ hybridization matrix between cluster and bath orbitals, whereas $\epsilon_{\a\b}$ is a $32\times32$ matrix of one-body terms within the bath, including possible superconducting pairing.
In principle, the matrix $\epsilon_{\a\b}$ could be diagonalized (this would change the values of the  hybridizations $\theta_{\mu\a}$), but we find it convenient and intuitive to allow pairing operators between bath orbitals.

The bath parameters are assumed to be spin independent, since we are not looking for magnetic ordering.
In order to probe superconductivity, we include singlet and triplet pairing operators within the bath.
Given two bath orbitals labeled by $r$ and $s$, the following pairing operators are defined:
\begin{align}
&~~\hat\Delta_{rs} = a_{r\up}a_{s\down}-a_{s\down}a_{r\up} &\mbox{(singlet)} \\
&\left.\begin{aligned}
\hat{d}_{rs}^{(x)} &= a_{r\up} a_{s\up} - a_{r\down} a_{s\down} \\ 
\hat{d}_{rs}^{(y)} &= i\left(a_{r\up} a_{s\up} + a_{r\down} a_{s\down}\right)  \\
\hat{d}_{rs}^{(z)} &= \left(a_{r\up} a_{s\down} +  a_{r\down} a_{s\up}\right) 
\end{aligned}\quad \right\}&\mbox{(triplet)}
\end{align}
In terms of the bath orbital numbering scheme defined on Fig.~\ref{fig:cluster_bath}b, the pairing terms in $H_{\rm imp}$ are
\begin{align}\label{eq:bath_params}
&~~~~ \Delta_1\left(\hat\Delta_{12} + \hat\Delta_{34} - \hat\Delta_{13} - \hat\Delta_{24}\right) \notag \\
&+ \Delta_2\left(\hat\Delta_{56} + \hat\Delta_{78} - \hat\Delta_{57} - \hat\Delta_{68}\right) \notag \\
&+ ip_{x1}\left(\hat d^{(x)}_{12} + \hat d^{(x)}_{34}\right) + ip_{y1}\left(\hat d^{(y)}_{13} + \hat d^{(y)}_{24}\right) \notag \\
&+ ip_{x2}\left(\hat d^{(x)}_{56} + \hat d^{(x)}_{78}\right) + ip_{y2}\left(\hat d^{(y)}_{57} + \hat d^{(y)}_{68}\right)
+ \mbox{H.c.}
\end{align}
These terms are included in the one-body matrix $\epsilon_{\a\b}$.

The AIM is characterized by 10 variational parameters, all illustrated on Fig.~\ref{fig:cluster_bath}b: bath energy levels $\epsilon_{1,2}$ (diagonal elements of $\epsilon_{\a\b}$), hybridization amplitudes $\theta_{1,2}$, singlet pairing amplitudes $\Delta_{1,2}$ and triplet pairing amplitudes $p_{x1}$, $p_{x2}$, $p_{y1}$, $p_{y2}$.
It turns out that, owing to the rather small value of $\lambda_{1,2}$ in Eq.~\eqref{eq:H0b}, the triplet bath parameters are too small to have an observable effect and can be neglected.
This reduces the number of independent bath parameters to six. 
The two clusters forming the supercell also happen to have the same bath parameter values in the converged solutions, which is expected from symmetry. 

For a given set of bath parameters, the AIM \eqref{eq:Himp} may be solved and the electron Green function computed.
The latter may be expressed as
\begin{equation}\label{eq:Gc}
\Gv_c(\omega)^{-1} = \omega - \tv_c - \Gammav_c(\omega) -\Sigmav_c(\omega)
\end{equation}
where $\tv_c$ is the one-body matrix in the cluster part of the impurity Hamiltonian $H_{\rm imp}$, $\Sigmav_c(\omega)$ is the associated self-energy, and $\Gammav_c(\omega)$ is the bath hybridization matrix:
\begin{equation}
\Gammav_c(\omega) = \thetav\frac1{\omega - \epsilonv}\thetav^\dg
\end{equation}
where $\thetav$ is the $16\times32$ matrix with components $\theta_{\mu\a}$ and $\epsilonv$ the $32\times32$ matrix with components $\epsilon_{\a\b}$.
Equation \eqref{eq:Gc} allows us to extract the cluster self-energy $\Sigmav_c(\omega)$ from computed quantities.
The fundamental approximation of CDMFT is to replace the full self-energy of the problem with the local self-energy $\Sigmav$.
More precisely, when the supercell contains more than one cluster, as is the case here, the supercell self-energy is the direct sum of the self-energies of the different clusters: $\Sigmav(\omega) = \bigoplus_c \Sigmav_c(\omega)$.
The lattice Green function is then approximated as
\begin{equation}\label{eq:Glatt}
\Gv(\kvt,\omega) = \frac1{\omega - \tv(\kvt) - \Sigmav(\omega)}
\end{equation}
where $\kvt$ is a wave vector of the reduced Brillouin zone (associated with the superlattice) and $\tv(\kvt)$ is the one-body Hamiltonian \eqref{eq:H0a}-\eqref{eq:H0c} expressed in that mixed basis of reduced wave vectors and supercell orbitals.
In our system, the matrix $\Gv(\kvt,\omega)$ has dimension $2\times 16=32$, because of the two clusters forming the supercell.

\begin{figure}[tp]
\includegraphics[width=0.98\columnwidth]{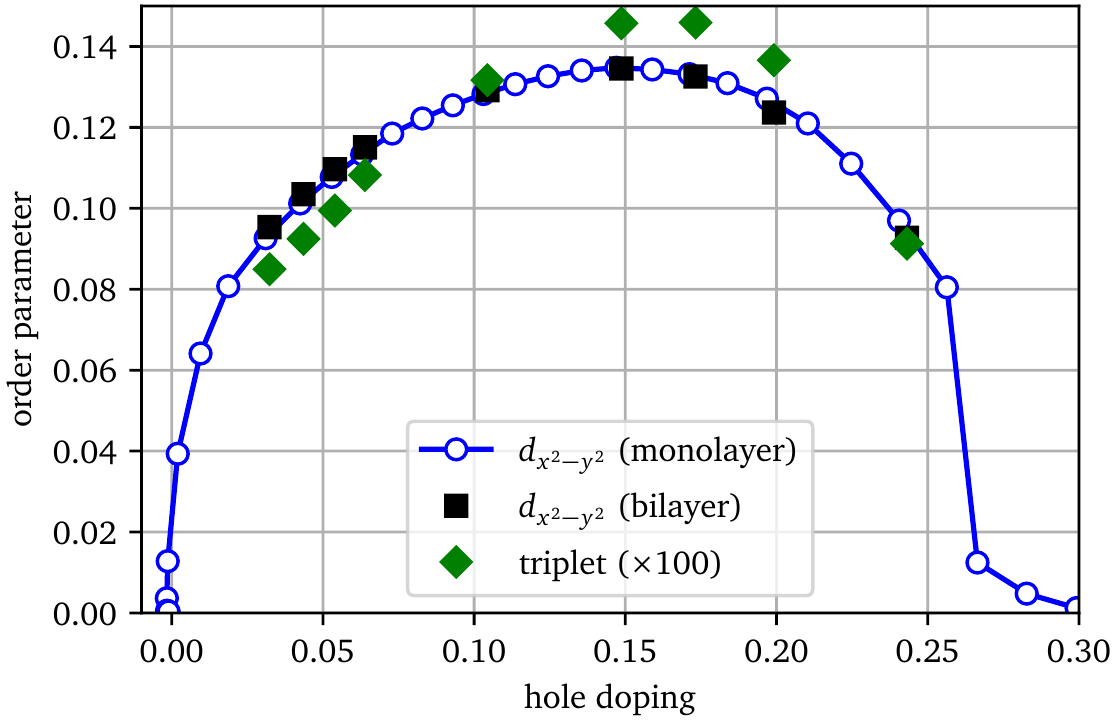}
\caption{(Color online) $d$-wave superconducting order parameter as a function of doping for $U=8$, in the single-layer model (open circles) and the bilayer model with spin-orbit coupling (black squares). The triplet component in the bilayer case is roughly 100 times smaller (green diamonds).}
\label{fig:OP}
\end{figure}

Let us finally summarize the self-consistent procedure used to set the bath parameters, as proposed initially in~\cite{ca.kr.94}: 
(i) trial values of the bath parameters are chosen on the first iteration. 
(ii) For each iteration, the AIM \eqref{eq:Himp} is solved, i.e., the cluster Green functions $\Gv_c(\omega)$ are computed using the Lanczos method, for each cluster.
(iii) The bath parameters are updated, by minimizing the distance function:
\begin{equation}
d(\epsilonv, \thetav) = \sum_{c, \omega_n} W(i\omega_n) \left[ \Gv_c(i\omega_n)^{-1} - \bar\Gv_c(i\omega_n)^{-1} \right]
\end{equation}
where $\bar\Gv_c(\omega)$ is the restriction to cluster $c$ of the projected Green function $\bar\Gv$, defined as
\begin{equation}\label{eq:GF}
\bar\Gv(\omega) = \int \frac{d^2k}{(2\pi)^2} \Gv(\kv,\omega)~~.
\end{equation}
(iv) We go back to step (ii) and iterate until the bath parameters or the bath hybridization functions $\Gammav_c(\omega)$ stop varying within some preset tolerance.

Ideally, $\bar\Gv_c(\omega)$ should coincide with the impurity Green function $\Gv_c(\omega)$, but the finite number of bath parameters does not allow for this correspondence at all frequencies. 
This is why a distance function $d(\epsilonv, \thetav)$ is defined, with emphasis on low frequencies along the imaginary axis.
The weight function $W(i\omega_n)$ is where the method has some arbitrariness; in this work $W(i\omega_n)$ is taken to be a constant for all Matsubara frequencies lower than a cutoff $\omega_c=2t$, with a fictitious temperature $\beta^{-1} = t/50$.  

The lattice Green function \eqref{eq:Glatt} can be used to compute the average of any one-body operator defined on the lattice.
In addition, we can go back to a fully wave vector-dependent representation $\mathcal{G}(\kv,\omega)$, where $\kv$ now belongs to the original Brillouin zone and $\mathcal{G}$ is a smaller, $8\times8$ matrix, by a procedure called \textit{periodization}.
The simplest periodization scheme is to Fourier transform $\Gv$ directly from the supercell to the original Brillouin zone, as follows~\cite{senechal2000a}:
\begin{equation}\label{eq:Gper}
\mathcal{G}_{ij}(\kv,\omega) = \frac1{N_c} \sum_{\rv, \rv'} e^{-i\kv\cdot(\rv-\rv')}G_{\rv i, \rv' j}(\kv, \omega)
\end{equation}
where $i,j$ are composite spin, Nambu and layer indices, and the difference between $\kv$ and $\kvt$ is an element $\Kv$ of the reciprocal superlattice: $\kv = \kvt+\Kv$. 
Note that $\Gv(\kv, \omega)=\Gv(\kvt, \omega)$ since $\Gv$ is by construction a periodic function of the reduced Brillouin zone.

\begin{figure}[tp]
\includegraphics[width=\columnwidth]{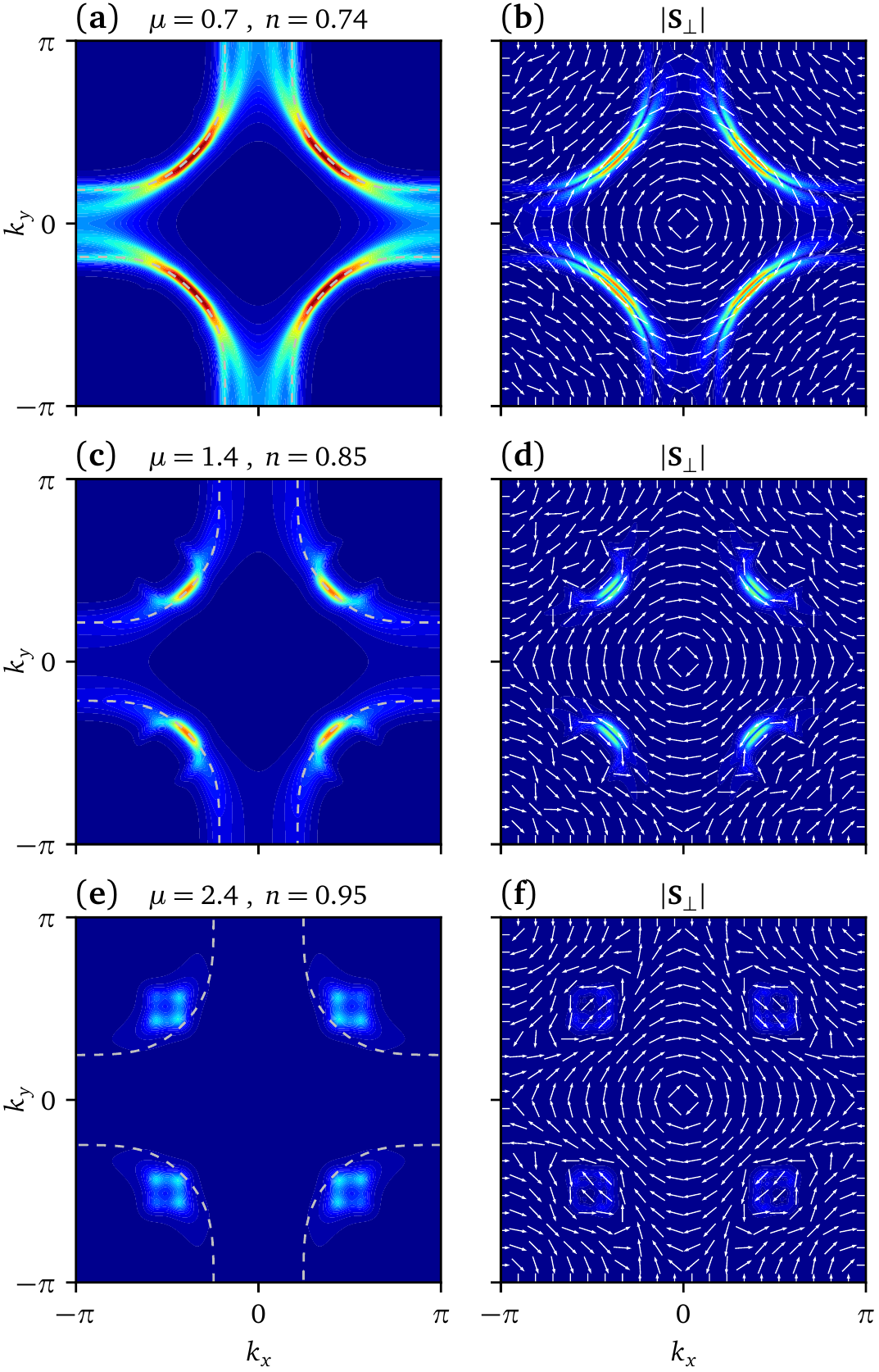}
\caption{(Color online) Spectral function and spin textures for $U=8$ and three values of electron density $n$. The left panels (a,c,e) show a color plot of the spectral function at the Fermi level within the Brillouin zone. The non-interacting Fermi surfaces for the same dopings are shown as a gray dashed line. The right panels (b, d, f) show the spin texture: the color plot is the magnitude of the transverse vector $\Sv_\bot(\kv) = (S_x(\kv,0), S_y(\kv,0))$ and the diluted set of arrows indicates its direction.
The color scale is the same for all three densities (blue is lowest, red is highest). Note that the four-dot structure at $n=0.95$ is an artefact of the cluster method, as explained in Ref.~\cite{verret2019}; only the dot touching the non-interacting Fermi surface is significant, the other three are ``harmonics''.}
\label{fig:texture}
\end{figure}

\begin{figure}[tp]
\includegraphics[width=0.75\columnwidth]{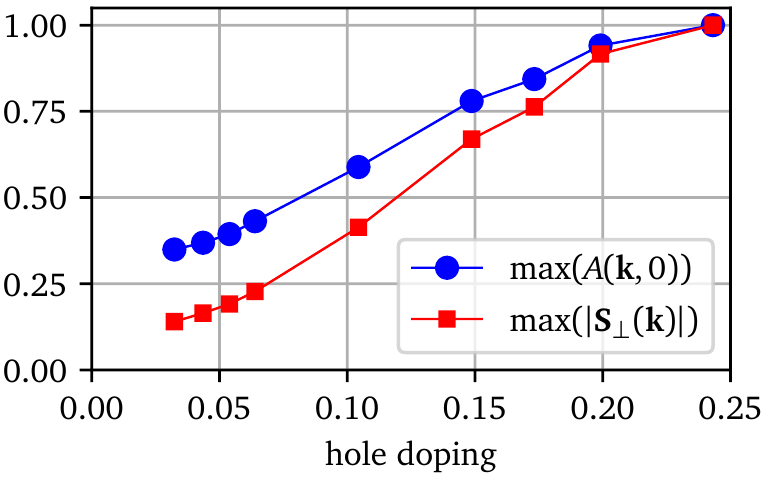}
\caption{(Color online) Maximum value of the spectral function $A(\kv,0)$ and of the spin texture $|\Sv_\bot(\kv)|$ over the Brillouin zone, for layer 1, as a function of doping, normalized to the same quantities at the last value of doping computed ($x=0.24$).
The drop as doping $x\to 0$ shows that the spin texture is more suppressed by the pseudogap physics than the spectral weight itself.}
\label{fig:wweights}
\end{figure}

\paragraph*{Results and discussion ---}

Figure \ref{fig:OP} shows the $d$-wave order parameter, computed from the Green function \eqref{eq:Glatt}, as a function of hole doping, for $U=8$.
This is the ground state average of the following operator:
\begin{equation}
\hat\Delta_{x^2-y^2} = c_{\rv,\up}c_{\rv+\xv,\down} - c_{\rv,\down}c_{\rv+\xv,\up} 
- c_{\rv,\up}c_{\rv+\yv,\down} + c_{\rv,\down}c_{\rv+\yv,\up} + \mathrm{H.c.}
\end{equation}
where $\xv$ and $\yv$ denote the nearest-neighbor vectors on the square lattice.
The blue curve is obtained in a single-layer model, without spin orbit coupling.
The black squares are obtained in the current bilayer model, and differ very little from the single layer values, because of the small value of both $t_z$ and $\lambda_{1,2}$.
The green diamonds are the average of the following triplet operator:
\begin{equation}
\hat d^{(y)}_{x} = c_{\rv,\up}c_{\rv+\xv,\up} + c_{\rv,\down}c_{\rv+\xv,\down} + \mathrm{H.c.}
\end{equation}
Note the factor of 100 in the scale.
A similar operator defined along the $y$ axis with the $x$ component of the triplet $d$-vector has equal expectation values.

The periodized Green function \eqref{eq:Gper} give us access to quantities observable by SARPES, such as the spectral function:
\begin{equation}\label{eq:A(k,w)}
A(\kv,\omega) = -\Im\Tr_N\Big(\bm{\mathcal{G}}(\kv,\omega)\Big)~~,
\end{equation}
where $\Tr_N$ means a trace over spin and layer indices that excludes the Nambu sector.
One can also extract spin information by projecting the Green function $\bm{\mathcal{G}}$ on various spin directions.
We thus define the following spin spectral functions:
\begin{equation}\label{eq:As(k,w)}
S_a(\kv,\omega) = -\Im\Tr_N\big(\sigma_a\bm{\mathcal{G}}(\kv,\omega)\Big)\qquad (a=x,y,z)
\end{equation}
One can also define the corresponding layer-resolved quantities.

Figure \ref{fig:texture} shows the spin texture measured in three of the superconducting solutions obtained, for three different values of density $n$ (underdoped, optimally doped and overdoped), on the first layer.
The left panel shows a color plot of the spectral function \eqref{eq:A(k,w)} at the Fermi level (the non-interacting Fermi surface is indicated by a dashed line).
The right panel shows the magnitude of the projection on the $x$-$y$ plane of the spin spectral function \eqref{eq:As(k,w)}, on which we superimposed a (diluted) vector plot indicating the direction of the spin in the $x$-$y$ plane, as given by the vector $\Sv_\bot(\kv)=(S_x(\kv,0),S_y(\kv,0))$. 
On this plot the arrows only provide a direction, not a magnitude.
The color scale is the same for all three values of electron density in the figure.
The structure of the spin texture is similar for all three cases illustrated: it is made of a clockwise rotating pattern around $\kv=(0,0)$ and around $\kv=(\pi,\pi)$, on the first layer.
The corresponding quantities on the second layer are obtained simply by reversing the arrows.

As expected, the amplitude of the spin texture is maximal in the vicinity of the non-interacting Fermi surface, with the important proviso that the pseudogap phenomenon suppresses the spectral weight (and the amplitude of the spin texture) away from the diagonals as we go in the underdoped regime.
Exactly on the Fermi surface, at least near the diagonals, the spin texture vanishes as it must change direction, leading to split maxima along the diagonals on the right panels of Fig.~\ref{fig:texture}.
Figure~\ref{fig:wweights} shows the drop of the spectral weight (indeed its maximum value over the Brillouin zone at the Fermi level) as doping $x$ decreases (blue circles).
This suppression is even more pronounced for the maximum value of $|\Sv_\bot|$ (red squares).

Overall, the known physics of the two-dimensional, one-band Hubbard model is not affected by the presence of the spin-orbit coupling between the two layers or by the inter-layer coupling.
The $d$-wave order parameter (Fig.~\ref{fig:OP}) is not affected in any visible way and the triplet component of the order parameter is two orders of magnitude smaller than the singlet component.
The spin-orbit coupling and the associated spin texture do not interfere with the pseudogap physics illustrated by the concentration of the spectral weight along the diagonal as one gets closer to half-filling.
The spin-orbit coupling is too small to make the system topological~\cite{na.ta.12}, and we checked that the Chern number, computed by adding the real part of the zero-frequency self-energy to the non-interacting Hamiltonian~\cite{wa.zh.12a, wa.zh.12b}, is indeed zero in all the cases studied.
The amplitude of the spin texture, however, is more suppressed in the underdoped region than the spectral function is (roughly twice as much).

\begin{acknowledgments}
Discussions with Yuehua Su and A.-M. Tremblay are gratefully acknowledged.
Computing resources were provided by Compute Canada and Calcul Qu\'ebec.
X.L. is supported by the National Natural Science Foundation of China (Grant No. 11974293) and the Fundamental Research Funds for
Central Universities (Grant No. 20720180015).
\end{acknowledgments}


%

\end{document}